# Effects of carbon nanotubes and graphene oxide absorbers on the noise of mode-locked fiber lasers

*Xiaohui Li[1], Kan Wu[1], Xuechao Yu[1], Yonggang Wang[2], Yishan Wang[2], Bo Meng[1], Yulong Tang[1], Xia Yu[3], Ying Zhang[3], Zhipei Sun[4], Perry Ping Shum[1,5], Qi Jie Wang[1, 5]\**

1. OPTIMUS, Centre of Excellence for Photonics, School of Electrical and Electronic Engineering, Nanyang Technological University, 50 Nanyang Ave., 639798, Singapore
2. State Key Laboratory of Transient Optics and Photonics, Xi'an Institute of Optics and Precision Mechanics, Chinese Academy of Sciences, Xi'an 710119, China
3. Singapore Institute of Manufacturing Technology, 71 Nanyang Drive, 638075 Singapore
4. Department of Micro- and Nanosciences, Aalto University, PO Box 13500, FI-00076 Aalto, Finland
5. Centre for Disruptive Photonic Technologies, Nanyang Technological University, 21 Nanyang Link, 63737, Singapore

*E-mail: qjwang@ntu.edu.sg

ABSTRACT: Phase noise is very important for the ultrafast pulse application in telecommunication, ultrafast diagnose, material science, and biology. In this paper, two types of carbon nano-materials, single-wall carbon nanotube and graphene oxide, are investigated for noise suppression in ultrafast photonics. Various properties of the wall-paper SAs, such as saturable intensity, optical absorption and degree of purity, are found to be key factors determining the phase noise of the ultrafast pulses. A reduced-noise femtosecond fiber laser is experimentally demonstrated by optimizing the above parameters of carbon material based SAs. The phase noise reduction more than 10 dB at 10 kHz can be obtained in the experiments. To our knowledge, this is the first time that the relationship between different carbon material based SAs and the phase noise of mode-locked lasers has been investigated. This work will pave the way to get a high-quality ultrashort pulse in passively mode-locked fiber lasers.

# 1. Introduction

Ultrashort optical pulses have been highly demanded in scientific studies and applications, e.g., high-speed optical laser sources for optical communication, optical frequency comb for optical metrology and time-resolved studies of ultrafast nonlinear phenomena, seeds sources for laser amplifiers and supercontinuum sources, as well as the optical nano-machining for the materials processing etc [1-10]. Among different techniques, mode-locked laser technology is the most efficient means that generates ultrafast optical pulses. Phase noise, timing jitter, and other noise characteristics of the mode-locked lasers are the important factors that should be considered in many applications such as long-range optical fiber communications, high-resolution optical sampling, and frequency comb. As a result, there has been growing interests in the development of low-noise, ultrashort-pulse laser.

Many saturable absorbers (SA) have been utilized to achieve mode locking, which can satisfy aforementioned requirements. Semiconductor saturable absorber mirrors (SESAMs) as one type of mode lockers are emerging as an enabling technology owing to their low noise and pulse-to-pulse optical phase-coherence [11,12]. However, the fabrication typically requires a complex and expensive manufacturing process. Therefore, novel SAs with better characteristics, lower cost, and easier integration are of great interests [13,14]. Carbon materials have excellent performance, which can be potentially utilized as SAs for the generation of ultrashort pulse [15-17]. Various carbon materials, such as single-wall carbon nanotube (SWNT), graphene, graphene oxide (GO), and even graphite, have attracted tremendous interests due to their unique physical properties. SWNT is a direct-bandgap material with its bandgap dependent on the diameter and chirality [15,18]. However, the fabricated nanotubes are often with hybrid chiralities, which lead to tremendous advantages in optoelectronics, such as large optical nonlinearity, ultrafast carrier relaxation time (sub-picosecond), environmental robustness, and low saturation intensity (SI) [13,14,19-25]. On the other hand, graphene, as a two dimensional allotrope of carbon atoms on a honeycomb lattice with conical band structure, could become the next disruptive technology [26]. Graphene has broad applications in photonics and optoelectronics, such as broadband saturable absorbers (SAs) [27-31], surface plasmonics [32], photodetectors [33,34], optical modulator [35], polarizers [36], and transparent electrodes [37]. While GO can be considered as the insulating and disordered analogue of the highly conducting crystalline graphene [38]. Experiments also show that few-layered GO also has ultrafast carrier relaxation and large optical nonlinearities [38,39], potential for ultrafast photonics [26,38,39]. All the carbon materials aforementioned have been demonstrated to have potential applications in various ultrafast lasers [13-15,26,31,40-42], (such as fiber [24,27,31], waveguide [43,44], solid-state [45,46], and semiconductor lasers [47,48]). Furthermore, these carbon materials exhibit variety of advantages [13,14], which lie in the advantage of broadband absorption [13,14], large optical nonlinearity [13,14], fast recovery time, easy to fabrication [49], and low cost [50]. Our prepared carbon materials wall paper absorbers (such as SWNT wall paper, GO wall paper as demonstrated here), fabricated by using vertical evaporation method, have been successfully used as SAs in the passively mode-locked lasers [42,51].

Controlling the noise characteristics (e.g., reducing timing jitter by 24% [52], and reducing phase noise by 34% [53]) has been studied in a passively mode-locked fiber laser based on SWNT [52,54,55] and graphene [53]. These studies show that SWNT and graphene play key roles in reducing noise of generated pules due to their saturable absorption [24]. However, various crucial factors of these SAs, such as operating wavelength, modulation depth (MD), and SI, which seriously influence the noise performance, have not yet been acknowledged. For a given passively mode-locked fiber lasers, the relationship of phase noise and SAs are still unknown at present. Another important aspect is that different carbon materials used for mode locking also need to be investigated, which have potential application in low-noise mode-locked fiber lasers.

In this paper, we investigate different carbon materials as SAs (such as SWNT, GO wall papers). The fabrication processes of different materials have been optimized based on their different physic-chemical characteristics. Various characterization methods (such as Raman spectra, linear absorption, and SEM) were utilized to improve the fabrication process (e.g., purification of the SAs). The results show that the SWNT induced larger loss than GO SA for each piece (PC). The GO SA shows fewer bubbles than SWNT SAs, which means that the GO SAs are purer than SWNT SAs. The phase noise performance of different wall paper SAs with different PCs is compared in a passively mode-locked Er-doped fiber laser, which operates in the anomalous dispersion regimes. We show that mode locking can be obtained by using different PCs of the absorber, respectively. The SWNT from 1 PC to 3 PCs can mode lock the fiber laser stably, so well as 1 PC to 4 PCs of the GO wall paper SA. The results show that the phase noise can be optimized by varying the number of the PCs of SA films, through controlling the cavity loss and the saturable intensity of the absorbers. More than 10-dB phase noise can be reduced by controlling the parameter of SAs. In addition, the purity of the SAs also influences the noise characteristics of the mode locking.

## 2. Characterization of the carbon material wall-paper SAs

We investigate different characteristics of the SWNT and GO SAs, which can potentially affect the phase noise of the passively mode-locked fiber lasers. The Raman spectra, linear transmission of the SA film, SEM of SWNT and GO wall papers, and nonlinear transmission of different PCs of the SA films are studied.

The fabrication of different carbon materials mainly utilizes the vertical evaporation method [42]. However, different SAs have different processing ways due to their characteristics. The detailed fabrication processing of SWNT SA and GO SA are shown in the Method part. In order to characterize the SWNT- and GO-polyvinyl alcohol (PVA) wall-paper samples, Raman spectroscopy is utilized to measure the "finger-prints". The spectrum of SWNT-PVA wall paper SA is shown in Fig. 1(a). The inset is the micrograph of SWNT-PVA films. The absorber was excited by a 532 nm lasers. Different parameters such as diameter, electrical characteristic, direction, and chirality are considered and can be derived from the main first- and second-order Raman modes of the Raman spectrum. Different bands represent different characterization of SWNT wall papers. Radial-breathing-mode (RBM) band (100-400 $cm^{-1}$) is induced from radial vibration of the carbon nanotube in phase, which can also be utilized to estimate the diameter of the tube. D band (1300-1400 $cm^{-1}$) is related with the breathing motions of the $sp^2$ carbon atoms in a ring. It reflects the defects on the nanotube surface. While G-band (1500-1600 $cm^{-1}$), usually consist of two sub-bands, is related with the axial and circumferential in-plane vibrations [56]. It can be seen clearly from the Raman spectrum which is the characteristic of SWNT with high intensity of G-band. The background as seen from the Raman spectrum is due to the impurity, the optical scattering, and the PVA induced additional background noise. Figure 1(b) is the Raman spectrum of the GO-PVA films. The inset is the micrograph of GO-PVA films. There are some characteristic bands of D, G as seen from Raman spectrum. Different peaks represent different physical meanings. D band at around 1366 $cm^{-1}$ is induced from the structural defects generated by attaching the hydroxyl and epoxide groups on the carbon basal plane. The G band at around 1610 cm$^{-1}$ is related to the first-order scattering of phonons of E$_{2g}$ symmetry [57]. There are some other peaks and the background as seen from the Raman spectra, which are related with the fabrication process induced by some polymeric materials in the SAs [58].

An UV-Visible-NIR spectrophotometer (Agilent Technologies, Cary 5000) was employed to measure the linear optical transmittance of the SWNT wall-paper absorber with different PCs as shown in Fig. 2(a). As the PC number increases, the transmittance will decrease. For example, the transmittance of single PC SWNT wall paper absorber is 78.23%, while it decreases to 39.23% for four-PC SWNT absorber at 1550 nm, which induces much linear loss if it is inserted in the laser cavity PC by PC. The linear optical transmittance of GO-PVA absorber film is shown in Fig. 2(b). The transmittance of single PC GO absorber around 1550 nm is 82.13 %. While increasing the PCs of the GO SAs, the transmittance becomes small. For a four-PC GO-PVA device, the transmittance reduces to 52.07 %, which will cause about 3-dB loss in the fiber cavity. SWNT films induce slightly larger loss than GO films for the same number of PCs. This is due to that high SWNT concentration is used, which can be controlled during the fabrication process.

We also check uniformity of different SA polymer films (i.e., SWNT- and GO-PVA films). Figures 3(a) and (b) show the scanning electron microscope (SEM) image of the cross-section of SWNT and GO wall papers. The average thickness of the SWNT SA is about 49.53 μm as show in Fig. 3(a). The SWNT-PVA is not pure at one side of the surface. Figure 3(b) shows the zoom-in image of the SWNT SA in a smaller range. We can see there are lots of small bubbles and large holes which significantly increase the impurity of the SAs. As a result, there are much optical scattering effects leading to loss in the fiber cavity when it used as SAs. Figure 3(c) and (d) show the SEM of the GO wall paper absorbers in different ranges. The average thickness of

GO wall paper absorber is 33.99 μm as shown in Fig. 3(c). The roughness of the facet is due to the cutting tool of the scissor. GO can be dispersed very well into water, so the fabrication processing is free of any surfactants such as sodium dodecyl sulfate (SDS), which can significantly reduce the non-saturable scattering losses of the SAs. As shown in Fig. 3(d), there are less air holes or bubbles in the GO-PVA polymer film, which shows the improved uniformity of the SA. The bright side is due to the less free electrons in this region during the measurement.

The nonlinear SA property of SWNT and GO SAs as a function of light intensity can be expressed as [59]

$$\alpha(I) = \frac{\alpha_S}{1 + I/I_s} + \alpha_{NS} \tag{1}$$

where $I_s$ is the SI, $\alpha_S$ and $\alpha_{NS}$ are the saturable and nonsaturable absorption, respectively.

Since the MD and SI are also very important for the pulse shaping and the noise properties, we also investigate the nonlinear transmission of different SA films with different PCs. The nonlinear transmission of the SA with different PCs is measured based on power-dependent measurements [59,60]. We use a commercial Er-doped fs fiber laser (MENDOCINO Femtosecond Fiber laser, CARMAR LASER Company) as a laser source, which has a repetition rate of 50 MHz, pulse duration of 80 fs, and maximum 80 mW average output power. The input power to the composite is varied by means of a variable optical attenuator. A reference signal is taken using an optical coupler (OC) to detect the incident power. Different PCs of the SWNT- and GO-PVA thin films, cut into many small (2 mm$^2$) composites, are directly sandwiched between fiber connectors. The performance of various SWNT- and GO-PVA SAs is shown in Figs. 4(a) and (b), respectively. The MD for 1, 2, and 3 PCs of SWNT-PVA wall paper are 2.54

%, 4.68 %, and 5.18 %, respectively. 3-PC SWNT-PVA SA will induce almost 44.34 % nonsaturable absorption loss in the fiber-based SA components. The MD of 1, 2, 3, and 4 PCs of GO-PVA wall paper are 0.58 %, 1.25 %, 1.83 %, and 1.96 %, respectively.

The measured nonlinear properties are summarized in Table 1. We can see that the single-PC SWNT-PVA and GO-PVA SA components have the SI of 9.00 and 0.205 $MW/cm^2$. With the increase of the number of PCs, the nonsaturable losses are increased linearly for the two sets of SAs. However, the saturation intensities decrease but not strictly linear decrease. The thickness of the SA films plays the predominant role for the nonsaturable losses. We attribute the variation of saturation intensities to the non-uniform of the SA films and the optical scattering effect. The modulation depths are also varied, with different number of the PCs. In general, SWNT-PVA SA fiber components have larger MD than GO-PVA SA fiber components, while the SI of the GO-PVA SA is lower than the one of SWNT-PVA SA. When we increase the input power that the power intensity reaches 10 $MW/cm^2$, it may near the damage threshold of the PVA. As a result, the saturation intensity will become low.

## 3. Setup of the passively mode-locked fiber lasers based on wall-paper SA

In the experiments, we use the passively mode-locked fiber laser cavity as a platform to achieve mode locking based on different carbon materials with different PCs. The schematic diagram of proposed passively mode-locked fiber laser is shown in Fig. 5. The fiber cavity consists of a section of single-mode fiber, a 0.8-m-long Erbium-doped fiber (EDF), with core-pumped through a wavelength-division multiplexer (WDM) by a 976-nm laser diode (LD). A polarization insensitive isolator (PI-ISO) and a squeezed-type polarization controller (PC) are used to stabilize the mode-locking. Light is extracted from the unidirectional cavity through a 10 % output coupler (PC). The carbon material wall papers (SWNT-, GO-PVA films) are inserted between two fiber connectors to incorporate into the fiber cavity. Different PCs of carbon material thin films are incorporated into the fiber cavity, respectively. The output performances are measured by different commercial equipments. An rf spectrum analyzer (Rohde & Schwarz FSUP26) together with a 2-GHz photodetector is used to characterize the phase-noise spectrum of the output mode-locked pulse trains. An autocorrelator and an optical spectral analyzer (OSA) are used to measure the pulse duration and output spectrum, respectively.

**4. Experimental result**

For each material, the output properties of laser are measured when the laser is fundamentally mode locked at the maximum pump power where no spurs are observed on the optical spectrum. We change the number of thin film PCs incorporated in the cavity for each material to see how the different concentration of the materials affects the laser operation (including laser noise properties).

The SWNT wall papers from 1 to 3 PCs are used in the fiber laser cavity to achieve mode locking successfully. The detailed experimental results based on SWNT wall paper are shown in Fig. 6. The spectra, phase noise, rf spectra for different PCs SWNT wall paper, and the pulse width for 3-PC SWNT absorber are shown in Figs. 6(a), (b), (c), and (d), respectively. The spectral widths increase from 2.51 to 4.53 nm, when the mode locking can be obtained from 1 PC to 3 PCs. On the contrary, the output power decreases from -2.59 dBm to -4.03 dBm, which is due to the large loss induced in fiber cavity with the increase of the number of the SA PCs. The pulse width of mode-locking by using 3 PCs SWNT SAs is about 827.8 fs ($sech^2$-profile is assumed). There are relaxation oscillations for the mode locking by using 1 PC SWNT SA as seen from the RF spectra, which locates at both side of the RF spectrum. For the case of two or three PCs of the SWNT SA, there is no relaxation oscillation as seen from Fig. 6(c). The large cavity loss, MI, and SI can change the phase noise performance.

GO wall paper can mode lock the fiber laser from 1 to 4 PCs. The mode-locked spectra, phase noises, and corresponding RF spectra for different PCs can be obtained from Fig. 7(a), (b), and (c), respectively. The autocorrelation trace of mode locking with 4-PC GO absorber is shown in Fig. 7(d). If a $sech^2$-profile is assumed, the pulse width is about 953.8 fs. There also exits relaxation frequency for one-PC GO wall paper, which is smaller comparing to the one of SWNT wall paper. The spectral width increase from 1.6 nm to 2.9 nm and the output power decrease from -5.24 dBm to -5.57 dBm with the increase of the PCs of GO SAs from 1 to 3 PCs. The minimum phase noises are -110.8 and -120.7 dBc/Hz at 1 and 10 kHz for the mode locking by using 2-PC SAs. The timing phase noises for 2-PC and 3-PC GO-PVA film are relative smaller than other two cases. Because there exits too much loss in the cavity due to the thickness of the SAs for the case of 4-PC GO SAs, the phase noise is become a little high. If we insert more PCs GO SA in the fiber cavity, mode locking can't be obtained due to the increased loss in the cavities.

The experimental results are summarized in Table 2. It can be found that for SWNT, more PCs lead to lower output power and wider optical spectrum, but also lead to lower timing phase noise and no relaxation oscillation. In the case of GO, the timing phase noise shows a minimum for 2-PC and 3-PC GO-PVA film. The optimization of the phase noise can be explained by two factors. One of the factors is that increase of PCs of SA materials lead to moderate MD and SI which can reduces the timing phase noise. The other is that more PCs also lead to higher loss due to the nonsaturable absorption and scattering loss, which obviously worsens the laser operation and increases the timing phase noise. Therefore, there is an optimum point between these two mechanisms when we increase the GO-PVA film PCs incorporated into the laser cavity.

The phase noises of SWNT SA at 1 kHz, have similar value with different PC of SAs. The PCs of the SWNT have little effect on the phase noise of mode-locked fiber lasers at low frequency region. But the phase noise of the fiber laser based on SWNT SA can be optimized from -117.8 dBc/Hz to -129.3 dBc/Hz, which is improved by 11.5 dBc/Hz at 10 kHz. The noises of the mode-locked fiber laser are optimized a lot. The PCs both affect the phase noise of the fiber laser based on GO SA at 1 kHz and 10 kHz. The phase noise can be optimized from -102 dBc/Hz to -110.8 dBc/Hz at 1 kHz, which is improved by 8.8 dBc/Hz. The phase noise can be optimized from -113.1 dBc/Hz to -120.7 dBc/Hz at 10 kHz, which is improved by 7.6 dBc/Hz. The GO SAs have a uniform effect on the phase noise of the fiber lasers.

Considering the measured optical properties of different SAs, these results can be appropriately explained. In the first place, SWNT SAs have large linear transmission loss than GO SAs at 1.5 μm, so 4-PC SWNT SA can't be responsible for mode-locking in the fiber laser systems at pump level. Secondly, the purity of GO film is better than SWNT film, which will reduce the scattering loss in the fiber cavity. Once more, the MD and SI of SWNT SA is larger than GO SA at different number of PCs, which makes the optimized SWNT SA based fiber laser have a better phase noise (-129.3 dBc/Hz at 10 kHz) than the one based on GO SA (-120.7 dBc/Hz at 10 kHz). There are complex balances of different parameters for controlling the phase noise of the fiber laser based on SAs, which can't be determined by only one parameter for the optimizing processing. In general, the SA film induced loss, degree of purification, SI, and MD, can all together affect the phase noise of the mode locked fiber laser. More PCs of SAs doesn't mean better phase noise, because there exit a complex relationships between the phase noise and the different parameters of SAs. We can utilize the SA characteristics to optimize the performance of the fiber laser system.

## 5. Conclusions

In summary, the method to achieve high quality ultrashort pulses in the mode-locked lasers is investigated. Different carbon materials (SWNT, GO) as SAs are evaluated systematically for application in low-noise passively mode-locked laser. The characteristics of the wall-paper SAs, such as transmission loss, SI, MD, and degree of purity, can be controlled by changing the PCs of the SA and the method. It shows that the PCs of the SAs influence the noise characteristics of mode locking. The phase noise can be reduced more than 10 dB at 10 kHz by SWNT-PVA and more than 8 dB by GO-PVA in the experiments. This work proposes a method to achieve high quality ultrashort pulses.

**Acknowledgements.** We would like to acknowledge financial support from A*STAR SERC grant (Grant Number 112-290-4018) and A*STAR SERC Advanced Optics in Engineering Programme (Grant no.: 122 360 0004). This work is also partially funded by the CAS/SAFEA International Partnership Program for Creative Research Teams. The authors specially thank Prof. Rüdiger Paschotta from RP Photonics Consulting GmbH for offering great helps. The authors would also like to acknowledge Youde Shen for measuring the SEM image of the material, Huanhuan Liu and Xiaonan Hu for some useful discussions.

**Key words:** single-wall carbon nanotube, graphene oxide, saturable absorber, ultrafast laser, phase noise, noise suppression, mode locking, Er-doped fiber laser.

Figure Caption:

Figure 1 The Raman spectrum of (a) SWNT-PVA and (b) GO-PVA wall paper absorbers excited by a 532 nm laser.

Figure 2 (a) Linear transmittance curves of SWNT-PVA and (b) GO-PVA wall-paper absorbers with different numbers of PCs. The blue, green, yellow, and red curves represent the 1, 2, 3, and 4 PCs of absorbers, respectively.

Figure 3 SEM of the cutting section of SWNT and GO wall papers in different scales. SWNT: (a) The image with low magnification, (b) the zoom-in image. GO: (c) The image with low magnification, (d) the zoom-in image.

Figure 4 The nonlinear transmittance with different numbers of PCs of (a) SWNT- and (b) GO-PVA SAs. The measurement wavelength is 1560 nm.

Figure 5 Schematic experimental setup of the passively mode-locked erbium-doped fiber laser based on SWNT- or GO-PVA SA.

Figure 6 Output characteristics of passively mode-locked fiber laser based on SWNT wall paper absorbers. (a) Mode-locked spectra of SWNT films with 1, 2, and 3 PCs, (b) corresponding phase noise spectra, (c) corresponding RF spectra, (d) the pulse width as seen from autocorrelator for the mode locking with 3-PC SWNT SAs.

Figure 7 Output characteristics of passively mode-locked fiber laser based on GO wall paper absorbers. (a) Mode-locked spectra of GO films with 1, 2, 3, and 4 PCs, (b) corresponding phase noise spectra, (c) corresponding RF spectra, (d) the pulse width as seen from autocorrelator for the mode locking with 4 PCs of GO SAs.

Table caption:

Table 1 Nonlinear optical properties of SWNT- and GO-PVA integrated fiber devices.

Table 2 Comparison of mode locking properties with different carbon materials.

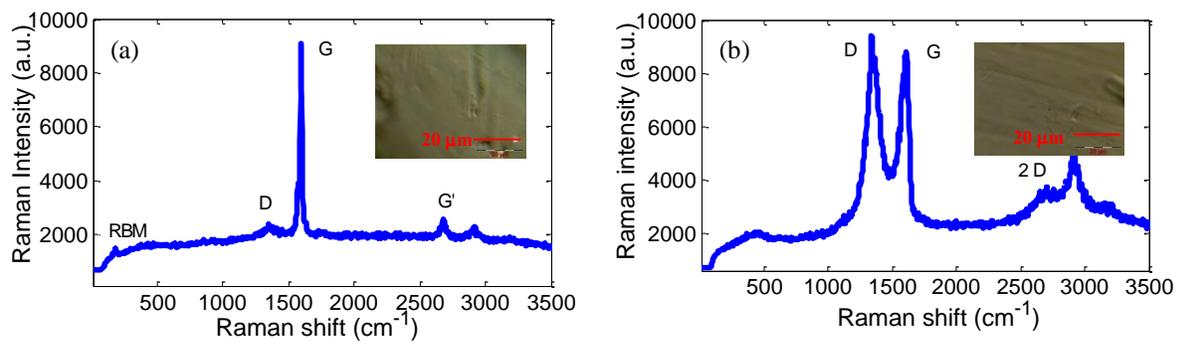

Figure 1 The Raman spectrum of (a) SWNT-PVA and (b) GO-PVA wall paper absorbers excited by a 532 nm laser.

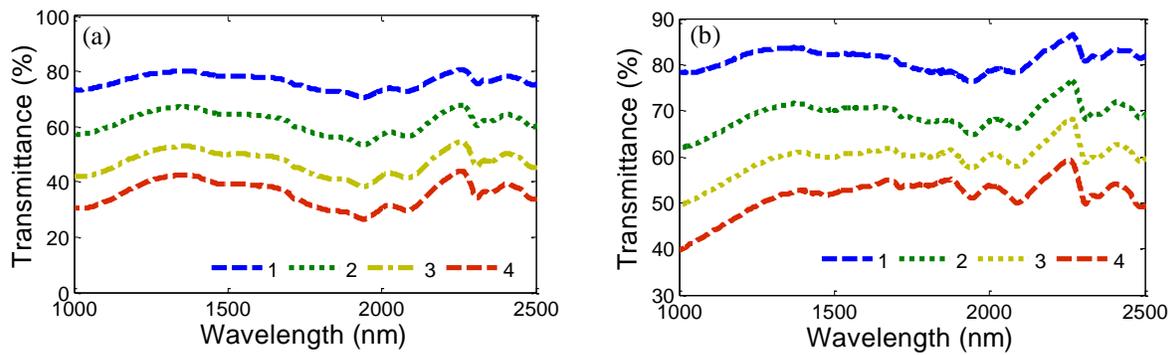

Figure 2 (a) Linear transmittance curves of SWNT-PVA and (b) GO-PVA wall-paper absorbers with different numbers of PCs. The blue, green, yellow, and red curves represent the 1, 2, 3, and 4 PCs of absorbers, respectively.

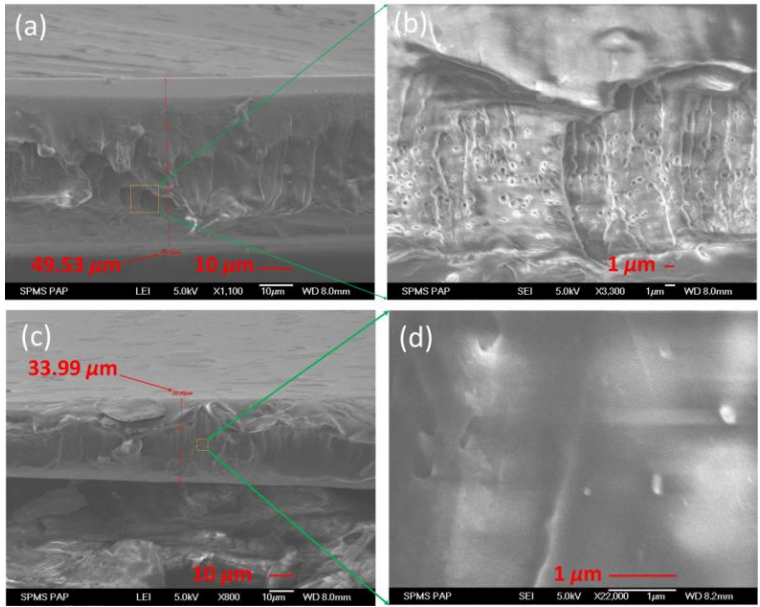

Figure 3 SEM of the cutting section of SWNT and GO wall papers in different scales. SWNT: (a) The image with low magnification, (b) the zoom-in image. GO: (c) The image with low magnification, (d) the zoom-in image.

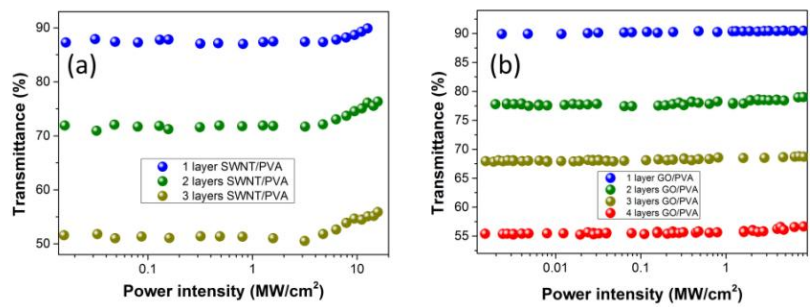

Figure 4 The nonlinear transmittance with different numbers of PCs of (a) SWNT- and (b) GO-PVA SAs. The measurement wavelength is 1560 nm.

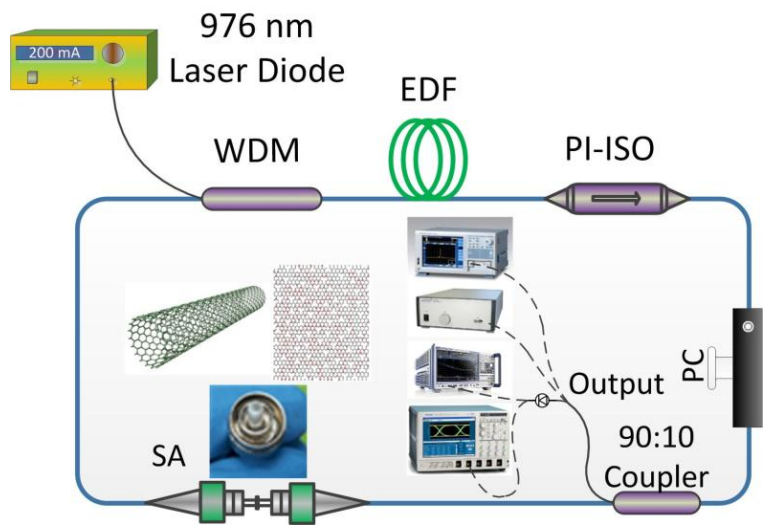

Figure 5 Schematic experimental setup of the passively mode-locked erbium-doped fiber laser based on SWNT- or GO-PVA SA.

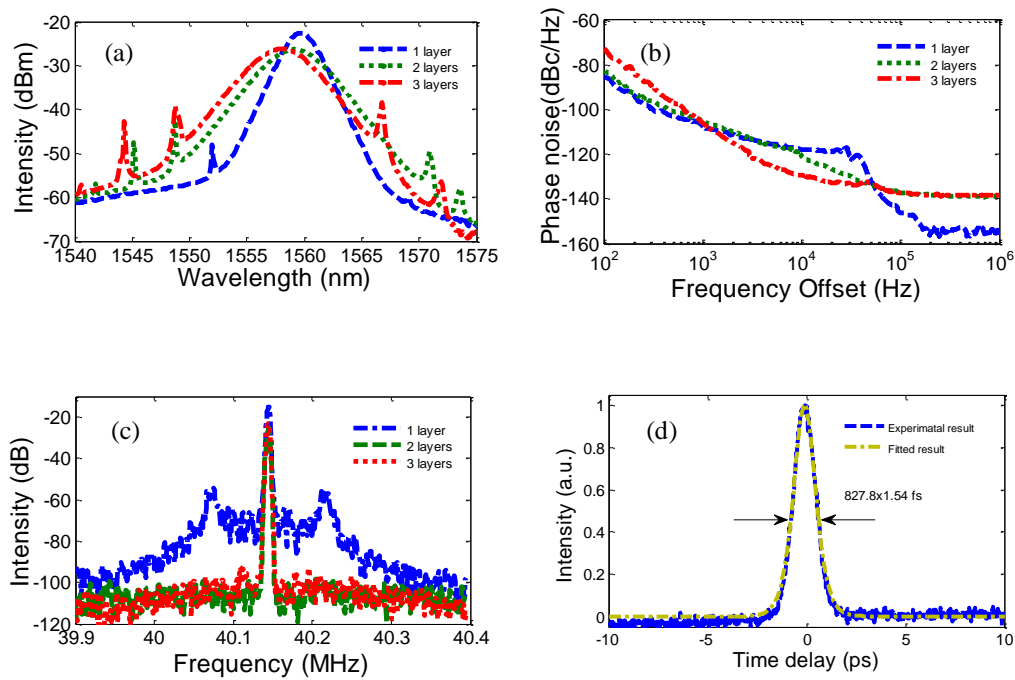

Figure 6 Output characteristics of passively mode-locked fiber laser based on SWNT wall paper absorbers. (a) Mode-locked spectra of SWNT films with 1, 2, and 3 PCs, (b) corresponding phase noise spectra, (c) corresponding RF spectra, (d) the pulse width as seen from autocorrelator for the mode locking with 3-PC SWNT SAs.

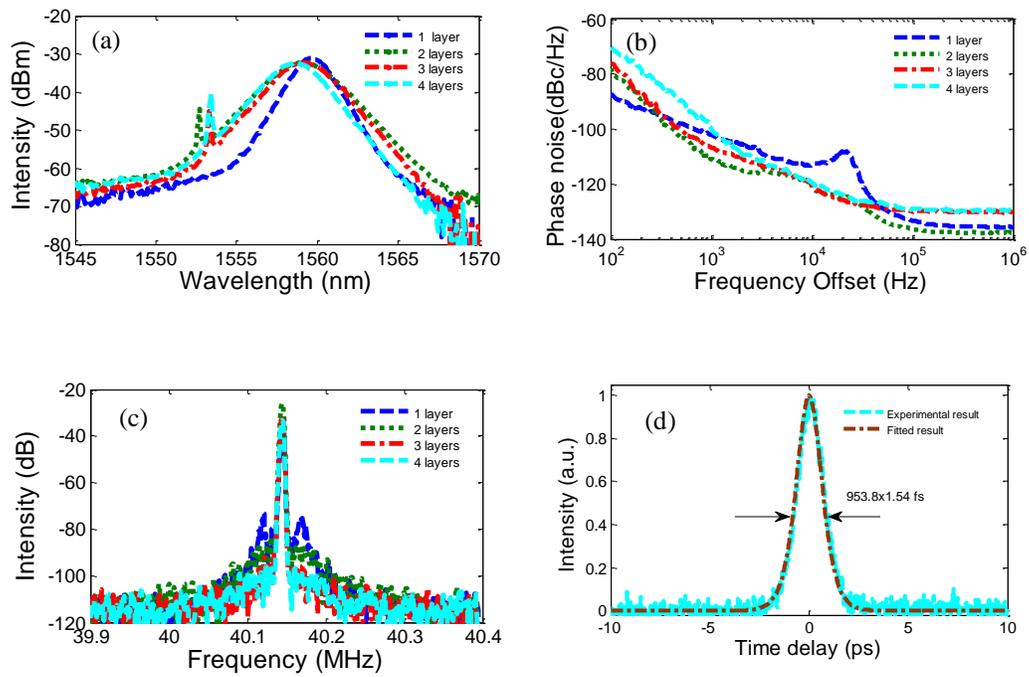

Figure 7 Output characteristics of passively mode-locked fiber laser based on GO wall paper absorbers. (a) Mode-locked spectra of GO films with 1, 2, 3, 4 PCs, (b) corresponding phase noise spectra, (c) corresponding RF spectra, (d) the pulse width as seen from autocorrelator for the mode locking with 4 PCs of GO SAs.

Table 1. Nonlinear optical properties of SWNT- and GO-PVA integrated fiber devices

| SAs | Numbers of PCs | Transmission | | Nonlinear transmission parameters | | Saturation intensity (MW/cm$^2$) |
|---|---|---|---|---|---|---|
| | | *Starting point (%)* | *Saturation (%)* | *Nonsaturable loss (%)* | *Modulation depth (%)* | |
| SWNT-PVA SAs | 1 | 87.38 | 89.92 | 10.08 | 2.54 | 9.00 |
| | 2 | 71.67 | 76.35 | 23.65 | 4.68 | 9.19 |
| | 3 | 50.48 | 55.66 | 44.34 | 5.18 | 9.66 |
| GO-PVA SAs | 1 | 89.91 | 90.49 | 9.51 | 0.58 | 0.205 |
| | 2 | 77.78 | 79.03 | 20.97 | 1.25 | 1.786 |
| | 3 | 67.95 | 68.78 | 31.22 | 1.83 | 1.820 |
| | 4 | 55.39 | 56.65 | 43.35 | 1.96 | 2.254 |

Table 2. Comparison of mode locking properties with different carbon materials.

|  | SWNT-PVA | | | GO-PVA | | | |
| --- | --- | --- | --- | --- | --- | --- | --- |
|  | *1 PC* | *2 PCs* | *3 PCs* | *1 PC* | *2 PCs* | *3 PCs* | *4 PCs* |
| **Pump power (mA)** | 207 | 234 | 293 | 156 | 168 | 184 | 203 |
| **Output power (dBm)** | -2.59 | -3.59 | -4.03 | -5.24 | -6.73 | -5.57 | -5.83 |
| **3-dB bandwidth (nm)** | 2.51 | 4.5 | 4.53 | 1.6 | 2.26 | 2.9 | 2.79 |
| **Center wavelength (nm)** | 1559 | 1559 | 1558 | 1559 | 1558.9 | 1559 | 1558 |
| **Output pulse energy (pJ)** | 13.7 | 10.9 | 9.8 | 7.45 | 5.29 | 6.9 | 6.5 |
| **RF carrier power (dBm)** | -19.75 | -25.56 | -23 | -28.1 | -26.2 | -32.5 | -33 |
| **Phase noise at 1kHz (dBc/Hz)** | -107.7 | -105.2 | -106 | -102 | -110.8 | -106.6 | -100 |
| **Phase noise at 10kHz (dBc/Hz)** | -117.8 | -120.7 | -129.3 | -113.1 | -120.7 | -120.6 | -119.4 |
| **Relaxation oscillation** | Yes | No | No | Yes | No | No | No |

X. H. Li, K. Wu, X. C. Yu et al.: Carbon based absorbers control the noise of mode-locked lasers

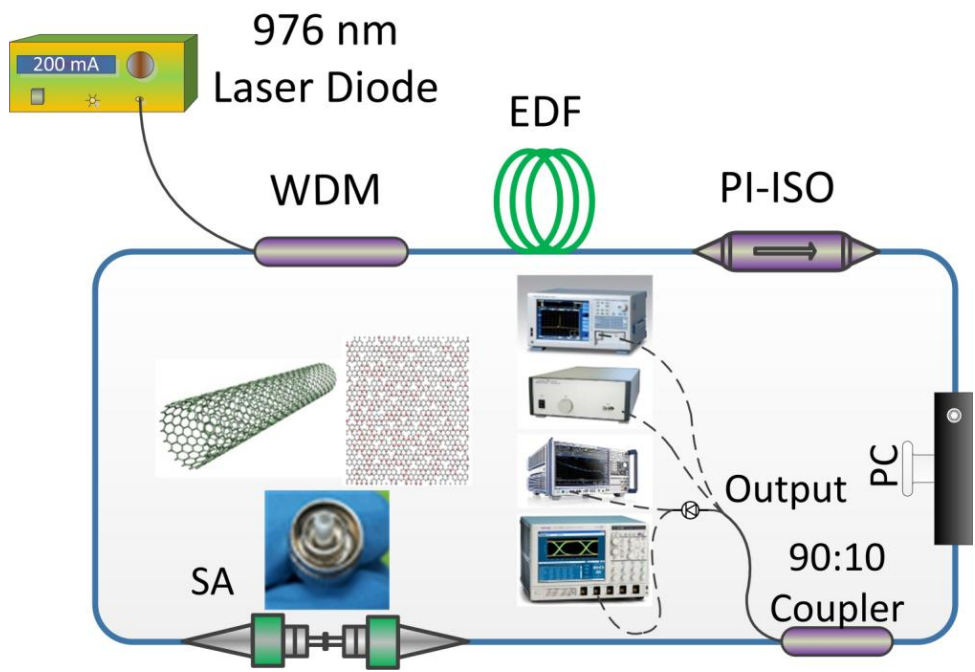

Eye-catching abstract figure